\documentstyle[preprint,aps]{revtex} 
\begin{document}
\draft
%\twocolumn[\hsize\textwidth\columnwidth\hsize\csname %
%twocolumnfalse\endcsname

\title{Ferromagnetism in 
the two dimensional $t$-$t^\prime$ Hubbard model at the Van Hove
density}
\author{ R. Hlubina$^1$\cite{byline}, S. Sorella$^1$, 
and F. Guinea$^2$} 
\address{$^1$ INFM, International School for Advanced Studies, 
Via Beirut 4, I-34014 Trieste, Italy\\
$^2$ Instituto de Ciencia de Materiales, Consejo Superior de 
Investigaciones Cient\'ificas, Cantoblanco, 28049 Madrid, Spain} 

\date{\today} 
\maketitle
\vskip -0.75 truecm
\begin{abstract}
\widetext {}Using an improved version of the projection quantum Monte
Carlo technique, we study the square-lattice Hubbard model with
nearest-neighbor hopping $t$ and next-nearest-neighbor hopping
$t^\prime$, by simulation of lattices with up to 20$\times$20 sites.
For a given $R=2t^\prime/t$, we consider that filling which leads to a
singular density of states of the noninteracting problem.  For 
repulsive interactions, we find an itinerant ferromagnet
(antiferromagnet) for $R=0.94$ ($R=0.2$). This is consistent with the
prediction of the $T$-matrix approximation, which sums the most
singular set of diagrams.
\end{abstract}
\pacs{ 75.10.Jm, 71.20.Ad, 71.27.+a}
%%%%%%%%%%%%%%%%%%
%]
%%%%%%%%%%%%%%%%%%

\widetext 

The understanding of itinerant ferromagnetism (FM) is a long-standing
problem of solid-state physics \cite{Vollhardt}. In search for a
generic model of FM, the Hubbard model, describing electrons from a
single band subject to a local electron-electron repulsion $U$, has
been investigated extensively. Motivated by an exact result of Nagaoka
\cite{Nagaoka}, most papers studied the stability of a fully polarized
state (believed to occur in the phase diagram of the Hubbard model at
large interaction strength and close to half filling) against single
spin flips. The results turned out to be strongly dependent on the
quasiparticle spectrum \cite{Hanisch}, and, generically,
unrealistically large $U$ were required to stabilize a fully polarized
state.  Motivated by the recent proofs of FM in certain models with
flat bands \cite{Mielke}, in this Letter we take a complementary
route, investigating the stability of the paramagnetic phase of a
model with a high density of states against FM at weak coupling.

We consider electrons on a square lattice with $L=l\times l$ (even
$l$) sites, described by the Hamiltonian
\begin{equation}
\label{hamiltonian}
H=-t\sum c^\dagger_{i,\sigma}c_{j,\sigma}
+t^\prime \sum c^\dagger_{i,\sigma}c_{j,\sigma} 
+U\sum_i n_{i,\uparrow}n_{i,\downarrow},
\end{equation}
where $t$ and $t^\prime$ are nearest and next-nearest neighbor
hoppings, respectively. The bare dispersion is $\varepsilon_{\bf
k}=-2t(\cos k_x+\cos k_y) +4t^\prime\cos k_x \cos k_y$ and its saddle
points are, for $0<R=2t^\prime/t<1$ studied here, at $(\pi,0)$ and
$(0,\pi)$. In what follows, we always consider the case when the
chemical potential is $-4t^\prime$, so that the noninteracting Fermi
surface crosses the saddle points and the noninteracting density of
states at the Fermi level diverges.  At this so-called Van Hove (VH)
density, both the particle-hole and particle-particle susceptibilities
$\chi_{\rm ph},\chi_{\rm pp}$ diverge at ${\bf q}=0$ and ${\bf Q}=
(\pi,\pi)$ and their singular parts are
\begin{eqnarray*}
\chi_{\rm ph}(0)&=&(1/2\pi^2t)(1/\sqrt{1-R^2})\ln(t/\omega),\\
\chi_{\rm ph}({\bf Q})&=&(1/2\pi^2t)\ln[(1+\sqrt{1-R^2})/R]\ln(t/\omega),\\
\chi_{\rm pp}(0)&=&(1/4\pi^2t)(1/\sqrt{1-R^2})\ln^2(t/\omega),\\
\chi_{\rm pp}({\bf Q})&=&(1/2\pi^2t)
[\arctan(R/\sqrt{1-R^2})/R]\ln(t/\omega),
\end{eqnarray*}
where $\omega$ is an infrared energy cutoff. It follows that, in the
mean field approximation, infinitesimal interactions cause symmetry
breaking: $U>0$ implies FM for $R>0.55$ and antiferromagnetism (AFM)
for $R<0.55$ \cite{Hirsch}, and $U<0$ leads to superconductivity
\cite{Scalapino}. The magnetic states are interesting because of their
metallic character.  However, particularly in the case of FM, it is
known that the use of a bare susceptibility in the Stoner criterion
leads to incorrect predictions. In order to go beyond mean field
theory, let us calculate the $U^2$ correction to the energy of a FM
state with magnetization $m=(N_\uparrow-N_\downarrow)/L$ where
$N_\sigma$ is the total number of electrons with spin
$\sigma=\uparrow,\downarrow$,
\begin{eqnarray*}
E^{(2)}=-{U^2\over 2}\sum_{\bf q}\int_{-\infty}^\infty
{d\omega\over 2\pi}\left[\chi_{\rm pp}({\bf q},i\omega)\right]^2.
%-U^2\sum {f_{{\bf k},\uparrow}f_{{\bf p},\downarrow}
%(1-f_{{\bf k^\prime},\uparrow})(1-f_{{\bf p^\prime},\downarrow})\over
%\varepsilon_{\bf k^\prime}+\varepsilon_{\bf p^\prime}-\varepsilon_{\bf
%k}-\varepsilon_{\bf p}}.
\end{eqnarray*}
$m\neq 0$ reduces the always negative $E^{(2)}$, because the anomalous
contributions of processes with momentum (energy) transfers ${\bf
q}\sim 0,{\bf Q}$ ($\omega\sim 0$) are cut off in the FM state.  We
have calculated the energy to order $U^2$ for $R=0.94$ on a
64$\times$64 lattice with 1612 electrons\cite{filling} for 10
closed-shell configurations with magnetizations $m<186/64^2$ and found
that the lowest possible $m=2/64^2$ was stable for all $U$.  Fig. 1
shows a similar result for a 16$\times$16 lattice. Thus a resummation
of an infinite set of diagrams is necessary.

In the following, we use the $T$-matrix approximation (TMA) in order
to go beyond perturbation theory. TMA is, in general, justified if the
interaction range is much shorter than interparticle spacing. This
situation is realized in the limit $R \rightarrow 1$.  Moreover,
scattering in the particle-particle channel is the most singular
process for all $R>0$. TMA has been shown previously to restrict
severely the possibility of low-density FM \cite{Mattis}. Following
Ref.\cite{Engelbrecht} we find that, in TMA, the spin-antisymmetric
Landau interaction function $f^a_{\bf k,k^\prime}= -U/[1+U\chi_{\rm
pp}({\bf k+k^\prime},\omega=0)]$. This means that the interaction
between those points of the Fermi surface which are in the
neighborhood of the VH points vanishes.  Assuming that the shape of
the noninteracting Fermi surface is not changed, let us consider now,
within Landau's Fermi liquid theory, the stability of the paramagnetic
state against Pomeranchuk instabilities. The change of the energy
under a local shift $\Delta_{{\bf k},\sigma}$ of the chemical
potential for spin-$\sigma$ electrons close to the point ${\bf k}$ on
the Fermi surface is
\begin{eqnarray}
\delta E&=&{1\over 2}\sum_\sigma\oint dk N_{\bf k} 
\Delta^2_{{\bf k},\sigma}\nonumber\\&+&
{1\over 2}\sum_{\sigma,\sigma^\prime}\oint dk N_{\bf k}\oint dk^\prime
N_{\bf k^\prime} f^{\sigma,\sigma^\prime}_{\bf k,k^\prime}
\Delta_{{\bf k},\sigma}\Delta_{{\bf k^\prime},\sigma^\prime},
\end{eqnarray}
where $N_{\bf k}\propto 1/v_{\bf k}$ is the angle-resolved density of
states, $v_{\bf k}$ is the Fermi velocity, and $\oint dk$ denotes an
integral along the Fermi surface.  For a ferromagnet $\Delta_{{\bf
k},\sigma}=\sigma\Delta$ and, consequently, $\delta E \propto
\Delta^2[1+F_0^a]$, where $F_0^a= \oint dk N_{\bf k}\oint dk^\prime
N_{\bf k^\prime} f^a_{\bf k,k^\prime}/\oint dk N_{\bf k}$. Let us
assume further that $v_{\bf k}\propto \delta k$, where $\delta k$ is
the distance between ${\bf k}$ and the closest VH point. For ${\bf
k,k^\prime}$ both in the neighborhood of the same VH point, we expect
$\chi_{\rm pp}({\bf k+k^\prime})\propto \ln^2[1/\max(\delta k,\delta
k^\prime)]$. The contribution of such ${\bf k,k^\prime}$ to the
integral for $F_0^a$ is $F_0^a\propto -\sqrt{U/t}$. Thus, intra-saddle
point processes do not lead to FM at weak coupling. However, if ${\bf
k,k^\prime}$ lie close to different VH points, we expect $\chi_{\rm
pp}({\bf k+k^\prime})\propto\ln[1/\max(\delta k,\delta
k^\prime)]$. This contributes $F_0^a\sim -(U/t)\int_{1/l}
dk/[k(1+A\ln(1/k))] \propto-\ln[A\ln l]$ where $A\propto U/t$ and $l$
is the system size. Thus the FM instability criterion $F_0^a<-1$ is
satisfied for arbitrarily small $U$ if $l\to\infty$. Note that TMA
predicts a much weaker divergence of $F_0^a$ with $l$, as compared to
the mean field prediction $F_0^a\propto -A\ln l$.  This is shown
explicitly for a 16$\times$16 lattice in Fig. 1.
%On the other hand,
%for $U<0$, the TMA reproduces the mean field prediction of
%superconductivity.  We conjecture that also for $U>0$ and $R\ll 1$,
%the mean field prediction of AFM is correct.

The above assumptions that the shape of the Fermi surface does not
change and that $v_{\bf k}\propto \delta k$ close to a VH point, are
supported by the Quantum Monte Carlo (QMC) data for $R=0.94$ and
$U/t=2$ on a $16\times 16$ lattice, which shows no appreciable change
of the Fermi surface. Moreover, our QMC results suggest that the
smearing of the momentum distribution function $n({\bf k})$ does not
increase close to the VH points. This is in contrast to the
predictions of the second-order perturbation theory in $U$ that the
quasiparticle residue $Z$ vanishes logarithmically at the VH points,
while being finite away from them.

Remarkably, for electrons described by $\varepsilon_{\bf k}=k_x k_y$
with the Fermi level $\varepsilon=0$, $Z$ is finite even at the VH
point in TMA.  Due to particle-hole symmetry, there is no change of
the Fermi surface for this model and the imaginary part of the
electron self-energy
%is \cite{Engelbrecht} $$ \Sigma^{\prime\prime}_{\bf k}(\varepsilon)=
%\sum_{\bf p}-\varepsilon<\varepsilon_{\bf p}<0
%[f_{\bf p}+b(\varepsilon+\varepsilon_{\bf p})] {\rm Im}\left[{U\over{
%1+U\chi_{\rm pp}({\bf k+p},\varepsilon+\varepsilon_{\bf p})}}\right],
%$$ where $b(\omega)$ is the Bose distribution function.  
at ${\bf k}=0$ is $\Sigma^{\prime\prime}(\varepsilon)\sim
-|\varepsilon|/\ln^2|\varepsilon|$ for $\varepsilon \to 0$. Thus the
wavefunction renormalization
$Z^{-1}=1-\partial\Sigma^\prime/\partial\omega|_{\omega=0}=
1-(2/\pi)\int_0^\infty d\varepsilon\Sigma^{\prime\prime}(\varepsilon)/
\varepsilon^2$ does not vanish, although the inverse lifetime differs
only marginally from the golden-rule result. Anyway, this anomalous
lifetime is not likely to appear in the transport properties of the VH
system.\cite{Hlubina}

The projection QMC method is obtained by filtering out the
ground-state component from a given trial state
$|\psi_0\rangle=\lim_{\tau \to \infty} e^{-H \tau } |\psi_T\rangle$.
Since the imaginary time propagation conserves the symmetries of the
trial function $|\psi_T\rangle$, it is also possible to study, e.g.,
different spin subspaces of $H$ by choosing a definite spin for
$|\psi_T\rangle$.  The many-body propagator $e^{-H \tau}$ is
implemented by use of the one-body time-dependent propagators
$U_\sigma (\tau,0)$ defined by the discrete Hubbard-Stratonovich
fields $\sigma (r,\tau)$ defined on each site $r$.  We use the usual
Trotter discretization of the total imaginary time $\beta$ for $0<\tau
< \beta$, namely $e^{-H \tau} =\sum_{\sigma} U_{\sigma}
(\tau,0)$.\cite{sorella} The ground state energy is obtained by $E_G
=\lim\limits_{\beta \to \infty} E_\beta$ where
$$
E_\beta= {\sum_\sigma \langle\psi_T|H U_\sigma (\beta,0)
|\psi_T\rangle\over
\sum_\sigma\langle\psi_T|U_\sigma(\beta,0)|\psi_T\rangle}.
$$
Analogously, all relevant correlation functions are obtained by Monte
Carlo sampling over the fields $\sigma (r,\tau)$, which become
computable provided the trial function $|\psi_T\rangle$ is chosen to
be a Slater determinant $|S\rangle$, as $U_\sigma(\tau,0)
|\psi_T\rangle$ remains a Slater determinant too.\cite{sorella}

The main improvements of the present scheme are :

1) Optimization of the trial function $|\psi_T\rangle$, which has been
extended to include not only simple Slater determinants $|S\rangle$
but also the more appropriate Gutzwiller wavefunction $|\psi_g\rangle=
e^{-g D } |S\rangle$, with a variational parameter $g$ controlling the
total number of doubly occupied sites $D$. This is obtained upon a
slight change of the one-body propagator $U_\sigma (\tau,0) \to
U_{\sigma,\sigma_g} (\tau,0)$, with extra discrete fields
$\sigma_g(r)$ needed to decouple the correlated part of the
wavefunction $e^{-g D}$ at the initial and final imaginary time
$\tau=0$ and $\tau=\beta$.

2) Use of the Gutzwiller wavefunction to implement an importance
sampling strategy to reduce statistical fluctuation of the ground
state energy, similar to what is already known for the Green function
QMC. \cite{ceperley} We define an estimator that satisfies the ``zero
variance principle'', namely that the variance of the energy is zero
if the trial wavefunction $|\psi_T\rangle$ approaches the ground state
$|\psi_0\rangle$. In order to satisfy this principle the energy has to be
computed at the initial and final imaginary times ($\tau=0,\beta$)
since only at those points the value of the estimator is independent
from the random fields $\sigma$ in the limit when $|\psi_T\rangle\to
|\psi_0\rangle$.  We define therefore
$$E_\beta ={ \sum_{\sigma,\sigma_g} E_{\sigma,\sigma_g}
w_{\sigma,\sigma_g} \over \sum_{\sigma,\sigma_g} s_{\sigma,\sigma_g}
w_{\sigma,\sigma_g} },$$ where
$w_{\sigma,\sigma_g}=|\langle S|U_{\sigma,\sigma_g}(\tau,0)|S\rangle|$
is the conventional weight (the sign being $s_{\sigma,\sigma_g}$). The
estimator is
\begin{equation} \label{estimator}
E_{\sigma,\sigma_g}= 
{\langle S| H_g U_{\sigma,\sigma_g}(\beta,0)  + U_{\sigma,\sigma_g}(\beta,0)
 H_{-g} |S\rangle\over 2 w_{\sigma,\sigma_g}},  
\end{equation}
and $H_g=e^{g D} H e^{-g D}$ is a nonunitary transformation of the
Hamiltonian, which can be computed by standard algebra.  Suppose now
that the ground state is well approximated by a Gutzwiller
wavefunction (the method can be clearly generalized to any Jastrow
wavefunction, $|\psi_g\rangle$=$e^{-g J} |S\rangle$, with $J$ any
two-body interaction term), then it is easy to show that the above
estimator will acquire a small variance, because the left eigenvector
of $H_g$ and the right one of $H_{-g}$ are well approximated by a
single Slater determinant $|S\rangle$.
 
In order to minimize the finite-size effects and also to stabilize the
simulation we restrict the Slater determinant to a product of
plane-wave Slater determinants in each spin sector $|S\rangle=
|S_\uparrow\rangle\bigotimes|S_\downarrow\rangle$, with the condition
to fill all degenerate single-particle levels in both spin sectors
(the closed-shell condition). In this case the total spin is $S=
|N_\uparrow -N_\downarrow|/2$.  Fig. 2a shows that the
improvement\cite{sornew} in the energy estimator allows to obtain
accurate values of energy in each spin sector even for small imaginary
time $\beta$, before the sign problem becomes serious. Among the
states obtained with a trial state $|\psi_T\rangle$ consistent with
the closed-shell condition, on a 16$\times$16 lattice the highest spin
sector has the lowest energy, as shown explicitly by comparing to the
lowest spin state (singlet in the thermodynamic sense). 
%This confirmes
%the FM instability previously obtained within TMA.\cite{Tremblay} 
We emphasize that, for a finite lattice, the instability occurs at a
finite coupling $U>U_c$, since the divergence of $F^a_0$ is cut-off in
this case. In fact, Fig. 2b shows that, for $U/t=4$ and $R=0.94$, the
singlet ground state is stable on the $10\times 10$ lattice, contrary
to the $16\times 16$ cluster. The tendency towards FM grows with
increasing cluster size also in the mean-field approximation, TMA, and
at the Gutzwiller variational level. The lattice-size dependence of
the difference between the energies of a fully polarized state, and of
a closed-shell state with minimal spin is shown in Fig. 2c.  Our QMC
data strongly support the FM state for $l\to \infty$.  The qualitative
agreement of TMA with the QMC data is striking. TMA tends to slightly
overestimate the FM correlations, but the improvement with respect to
the mean-field approximation and the Gutzwiller wavefunction (see
Fig. 1) is evident.

In order to confirm the prediction of AFM and superconductivity in the
remaining part of the phase diagram, we have studied the imaginary
time dependence of correlation functions in the following form:
\begin{equation} 
\label{timed} 
O(\tau)= {\langle\psi_T |e^{- \tau H }\hat O e^{-(\beta-\tau)H} 
|\psi_T\rangle
\over \langle\psi_T | e^{- \beta H } |\psi_T\rangle} .
\end{equation}  
For $\beta \to \infty$, $O(0)$ tends to the so called mixed average
$O(0)= {\langle\psi_T|\hat O |\psi_0\rangle\over\langle\psi_T|\psi_0
\rangle}$, which is a property of both the ground state and the trial
wavefunction itself.  Instead, for $\tau\sim\beta/2$, one obtains the
desired ground-state expectation value of $\hat O$.  Thus, if we
choose as a trial wavefunction the singlet paramagnetic Gutzwiller
wavefunction, i.e. a state without BCS or AFM order, an increase of
the corresponding correlations is expected
%(e.g. $O=  S(\pi,\pi)$, the spin-isotropic magnetic structure factor)
by varying $\tau$ from $0$ to $\beta/2$, whenever the electron system
truly supports some kind of order. The time-dependent functions
$O(\tau)$ defined in (\ref{timed}), directly indicate enhanced or
depressed correlations with respect to a reference finite-size
Gutzwiller state $|\psi_T\rangle$.  They represent therefore clear
fingerprints of the nature of the ground state $|\psi_0\rangle$ even
with a short imaginary-time propagation $\beta$, provided the
reference state $|\psi_T\rangle$ is sufficiently close to
$|\psi_0\rangle$.  A similar idea was used also in Ref. \cite{Heeb}.

In Fig. 3a we plot the spin-isotropic magnetic structure factor for
the AFM wavevector $Q=(\pi,\pi)$ as a function of the lattice size
$l$.  Only for $l>12$ the AFM correlations are clearly enhanced,
supporting the existence of a true AFM long range order.
Unfortunately, the need for $\beta t>2$ at large $l$ precludes the
possibility of a detailed finite-size scaling of AFM order.  For
negative $U$ (Fig. 3b) $s$-wave superconductivity with an order
parameter $s= { 1\over L} \sum_{k} c^{\dag}_{k,\uparrow}
c^{\dag}_{-k,\downarrow}$ should take place.  The plot of $O(\tau)$
where $\hat O=s^{\dag} s$ clearly shows this tendency even for a short
imaginary-time propagation,
%(which in this
%case can be made much longer because there is no sign problem)
in agreement with Ref. \cite{Scalapino}.

Based on the results of QMC simulations and on the analysis of TMA we
conclude that, in the $t$-$t^\prime$ Hubbard model at the VH density,
interactions tend to remove the singular density of states, via a
splitting of the Fermi surface (FM) or the opening of a gap at the VH
points (AFM and superconductivity).  This has been shown numerically
at realistic coupling strengths $U$ and is expected analytically for
all $U$, if we neglect the change of the quasiparticle
dispersion. Since we find magnetism at $U>0$ despite the finite-size
cut-offs, we expect that it appears, for $U$ large enough, in a finite
window of densities around the VH filling.  Our results for repulsive
interactions reduce in the limits $R\to 0$ and $R\to 1$ to AFM of the
half-filled $t^\prime=0$ Hubbard model and to an analogue of the
flat-band FM \cite{Mielke}, respectively. An interesting question
arises about the transition from AFM to FM as $R$ increases.  Since
already on the mean field level both FM and AFM susceptibilities are
suppressed for $R\approx 0.55$ as compared to their values at $R=0.94$
and $R=0.2$, respectively, a numerical solution of this question will
require even bigger lattices than those used here. We cannot exclude
that close to $R\approx 0.55$, there is no magnetic ordering and a
superconducting instability develops via the Kohn-Luttinger effect
\cite{KL}.

We thank J. Gonz\'alez, A.  Alvarez, M.  Fabrizio, G. Santoro, and
E. Tosatti for useful discussions. SS also thanks the CSIC (Spain) for
hospitality. The largest simulations were done on an IBM SP2 16-nodes
parallel machine of SISSA.  This work was supported in part by the EEC
Contract ERB-CHR-XCT 940438.

\begin{figure} \caption{
The energy per site (in units of $t$) as a function of $m$ for
$N=N_\uparrow+N_\downarrow=108$ electrons.\protect\cite{filling} Full
lines: free electrons (bars), mean-field approximation (triangles),
2$^{\rm nd}$ order perturbation theory (squares), and $T$-matrix
approximation without self-consistency condition for the electron
Green's function (circles). Dashed lines: optimal Gutzwiller
wavefunction (squares) and QMC data at $\beta t=2$ (triangles). }
\label{fig1}
\end{figure}
\begin{figure} \caption{ 
The energy per site (in units of $t$) 
(a) of a $L=16 \times 16$ lattice with $N=108$\protect\cite{filling}
 electrons as a function of $\protect\beta$, for $N_\uparrow=55$
 (total spin $S=1$, continuous line) and the $\beta$ converged one for
 the FM state (total spin $S=53$, dashed line);
(b) as a function of $m$ at $\protect\beta t= 2$ for $N=108$, $L=16
 \times 16$ (upper curve) and $N=44$, $L=10 \times 10$ (lower curve).
 The optimal $g=\protect\sim 0.65$ (see text) is used in all
 simulations. The error due to the imaginary time discretization
 ($\protect\Delta \protect\beta t= 0.2$) is negligible, as the new
 estimator (\protect\ref{estimator}) considerably reduces both this
 ($\protect\sim 10 \protect\div 100$ times) and the statistical error
 ($3 \protect\div 4$ times) at the optimal $g$.  
(c) Difference between the energy per site (in units of $t$) of
 the minimal-spin ($S=1$) state and of the fully polarized state, both 
 at the VH density, as a function of the lattice size.} 
\label{fig2}
\end{figure}
\begin{figure} \caption{ 
QMC simulation for $t^\prime/t=0.1$ at the VH density in the singlet
sector. (a) Isotropic magnetic structure factor $S(\pi,\pi)= { 1\over
L} \sum_{i,j} < \vec S_i \cdot \vec S_j > e^{i Q (R_i-R_j)}$ at the AF
wavevector $Q$ as a function of $\tau$ (see text) for $U/t=4$. The
curves (guides to the eye) correspond to square lattices $l \times l$
with $l=6,8,10,12,16,20$ ($N_\sigma=15,27,43,63,115,183$) from bottom
to top. The optimal Gutzwiller parameter $g=0.625$. (b) The same plot
as in (a) for the $s$-wave BCS order parameter described in the text
on a lattice $16\times 16$ for $U/t=-2$. The optimal $g=-0.3$ and
$\Delta \beta t=0.1$.}
\label{fig3}
\end{figure}

\begin{references}
\bibitem{filling} Here the VH density is obtained on a finite
lattice by filling the highest-energy wavevectors $(\pi,0)$ and
$(0,\pi)$ by two up-spin and no down-spin electrons. At such filling,
there is a large number of low-magnetization closed-shell
configurations.
\end{references}
\end{document}